\documentclass[nonacm]{acmart}

\usepackage{hyperref}
\usepackage{hyperxmp}

\AtBeginDocument{%
  \providecommand\BibTeX{{%
    \normalfont B\kern-0.5em{\scshape i\kern-0.25em b}\kern-0.8em\TeX}}}

\setcopyright{acmcopyright}
\copyrightyear{2018}
\acmYear{2018}
\acmDOI{XXXXXXX.XXXXXXX}

\acmConference[Conference acronym 'XX]{Make sure to enter the correct
  conference title from your rights confirmation emai}{June 03--05,
  2018}{Woodstock, NY}
\acmPrice{15.00}
\acmISBN{978-1-4503-XXXX-X/18/06}

\begin{document}

\title[Conversational AI as a Coding Assistant]{Conversational AI as a Coding Assistant: Understanding Programmers' Interactions with and Expectations from Large Language Models for Coding}


\author{Mehmet Akhoroz}
\email{akhoroz.m@northeastern.edu}
\affiliation{%
  \institution{Khoury College of Computer Sciences, Northeastern University}
  \city{Boston}
  \state{MA}
  \country{USA}
}

\author{Caglar Yildirim}
\email{c.yildirim@northeastern.edu}
\affiliation{%
  \institution{Khoury College of Computer Sciences, Northeastern University}
  \city{Boston}
  \state{MA}
  \country{USA}
}


\begin{abstract}

Conversational AI interfaces powered by large language models (LLMs) are increasingly used as coding assistants. However, questions remain about how programmers interact with LLM-based conversational agents, the challenges they encounter, and the factors influencing adoption. This study investigates programmers' usage patterns, perceptions, and interaction strategies when engaging with LLM-driven coding assistants. Through a survey, participants reported both the benefits, such as efficiency and clarity of explanations, and the limitations, including inaccuracies, lack of contextual awareness, and concerns about over-reliance. Notably, some programmers actively avoid LLMs due to a preference for independent learning, distrust in AI-generated code, and ethical considerations. Based on our findings, we propose design guidelines for improving conversational coding assistants, emphasizing context retention, transparency, multimodal support, and adaptability to user preferences. These insights contribute to the broader understanding of how LLM-based conversational agents can be effectively integrated into software development workflows while addressing adoption barriers and enhancing usability.
\end{abstract}






\maketitle

\section{Introduction}

Large language models (LLMs) such as ChatGPT, Claude, and GitHub Copilot have rapidly emerged as powerful conversational coding assistants, enabling developers and programmers to engage in natural language dialogues for code generation, debugging, and learning. Unlike traditional programming support resources such as static documentation, online Q\&A forums, and video tutorials, LLM-based conversational user interfaces (CUIs) facilitate interactive, iterative problem-solving. Users can refine prompts, request clarifications, and receive code suggestions tailored to their specific contexts, making these tools highly attractive for both students and professional developers \cite{Barke2023, Vaithilingam2022}.

The increasing adoption of LLM-based coding assistants raises several key questions: How do users engage with these tools? What aspects of LLM assistance are perceived as most effective or limiting? How do conversational strategies impact the quality of generated code? Prior research has begun addressing these topics, investigating how programmers integrate LLMs into their workflows, which features they use most, and how AI-generated code compares to conventional resources such as Stack Overflow and official documentation \cite{Liu2023ChatAssistComparison, Jiang2022}. 

Existing studies have also explored how programmers interact with LLMs, revealing both their strengths and limitations. Developers often use LLMs for code completion, error debugging, and API usage clarification, benefiting from their ability to generate syntactically correct and contextually relevant responses \cite{Ross2023, Nam2024}. At the same time, issues such as incorrect code generation, lack of contextual awareness, and security vulnerabilities have raised concerns about their reliability and the strategies users employ to validate outputs \cite{Sandoval2023, Vaithilingam2022}. In educational contexts, students have reported increased productivity and learning gains when using LLMs, yet concerns about over-reliance and reduced problem-solving skills persist \cite{Huang2024, Kazemitabaar2023a}.

In this paper, building on existing work we conducted an empirical study on how software developers use LLMs for coding tasks. Using a survey-based approach, we examined patterns of LLM use, user perceptions of effectiveness, and common interaction strategies. We also explored the strengths and limitations of LLM-based CUIs compared to traditional programming support tools. The results contribute to a growing body of research on AI-assisted coding, providing insights into how conversational programming interfaces influence coding workflows, learning processes, and productivity. Specifically, this study makes three key contributions to CUIs for LLM-based coding assistance:

\begin{enumerate}
    \item \textbf{Empirical evidence of LLM usage patterns, expectations, and barriers to adoption:} Our study provides empirical data on how students and developers use LLM-based assistants, highlighting a shift towards debugging support and concept clarification rather than direct code generation. In addition, we identify the primary reasons why some programmers choose not to use LLMs, including concerns about learning efficacy, trust and reliability, availability of alternative resources, and ethical considerations. These insights contribute to a broader understanding of the factors influencing LLM adoption in software development.
    
    \item \textbf{Identification of effective interaction strategies:} We extend prior research on prompt engineering and LLM validation strategies by identifying how users iteratively refine queries, request explanations, and compare multiple LLM-generated outputs to improve coding outcomes. Our findings suggest that programmers do not passively accept AI-generated responses but instead engage in adaptive query refinement and verification practices to maximize the utility of LLM-based assistants.
    
    \item \textbf{Design guidelines for custom LLM-driven coding assistants:} Based on user expectations and the identified barriers to adoption, we propose a set of design principles to guide the development of more effective LLM-integrated coding assistants. These guidelines emphasize context awareness, transparency, multimodal support, IDE integration, and adaptability to user preferences, ensuring that AI-powered tools complement existing workflows while addressing concerns about learning, trust, and ethical considerations.
\end{enumerate}

\section{Related Work}

Research on AI-assisted programming has examined how LLMs impact developer productivity, learning experiences, and coding practices. Several studies have assessed the usability of LLM-based coding assistants, identifying key interaction patterns and user behaviors.

\subsection{LLM Usage Patterns and Common Use Cases in Coding}
Previous studies indicate that both novice programmers (e.g. students) and experienced developers are turning to LLM-based assistants for a range of coding tasks. One common use case is code generation and synthesis. LLMs can generate boilerplate code and implementations for given specifications. For example, users often employ Copilot or ChatGPT to quickly produce a function or algorithm given a natural-language prompt \cite{Jiang2022}. This is especially handy for writing repetitive code or scaffolding (e.g. creating template code for a UI or data structure). Jiang et al. \cite{Jiang2022} found that participants likened LLM prompts to a new kind of programming “syntax,” using natural language to generate HTML, JavaScript, or other code from scratch.

Another common use case is API usage and lookup. Developers use LLMs as an intelligent lookup tool for libraries and APIs \cite{Jiang2022, tao2024harnessing, basu2024api}. Instead of searching documentation, users can ask the LLM how to accomplish a task with a certain library and often receive example code. Previous research showed that participants leveraged LLM assistants as a quick way to get the equivalent of an API example or “recipe,” accelerating tasks that would normally require combing through docs \cite{Jiang2022}.

Debugging assistance is yet another common use case \cite{zhang2023survey, etsenake2024understanding}. Users can paste error messages or buggy code into a chat and ask for help. LLMs often suggest fixes or point out likely causes of errors. However, as discussed later, the effectiveness here varies (e.g. ChatGPT may help with certain bugs but can be less reliable than searching specific error threads \cite{Liu2023ChatAssistComparison, etsenake2024understanding}
). Liu et al. \cite{Liu2023ChatAssistComparison} found that while ChatGPT could quickly propose solutions for algorithmic or library issues, traditional Q\&A forums like Stack Overflow were sometimes better at solving tricky bugs or exceptions.


\subsection{Perceived Usefulness and Effectiveness of LLM Assistance}
Overall, users perceive LLM-based coding assistants as highly useful, albeit with some caveats \cite{rasnayaka2024empirical}. Previous studies have found that programmers enjoy using these tools and often prefer having them available during coding \cite{nam2024using, Kazemitabaar2023}. For example, in a user study of GitHub Copilot (an LLM plugin for code editors), Vaithilingam \cite{Vaithilingam2022} found that the majority of the participants indicated a preference for using Copilot in everyday coding tasks because it provided a helpful starting point for implementations and reduced the need to search online and/or look through documentation. Even when Copilot did not outright solve a task, it gave users a chunk of code or an approach that they could then refine. This sentiment is echoed in other studies: developers report that tools like Copilot make coding more enjoyable and can reduce “blank page” paralysis by offering immediate suggestions \cite{bird2022taking, nizamudeen2024investigating, sobania2022choose, wermelinger2023using}.

While LLMs do not always result in more effective and efficient coding \cite{becker2023programming}, many users still consider LLM-assisted coding effective in a qualitative sense: it reduces mental effort and provides confidence or inspiration. Copilot users in one study felt it “significantly improved [their] programming productivity” subjectively, even when the measured speed increase was small \cite{Vaithilingam2022}. In industry settings, studies have reported developers feeling more productive and satisfied with their work when using AI coding assistants (e.g., enjoying coding more and experiencing less frustration) \cite{weber2024significant, weisz2024examining}. This indicates that perceived effectiveness (in terms of workflow satisfaction and reduced cognitive load) may be high, even if quantitative effectiveness (in terms of speed or correctness) is moderate. One explanation is that LLMs help users arrive at a solution with less mental strain, which users value highly, especially in creative or complex tasks. 

 In summary, developers generally perceive conversational LLMs as useful assistants that enhance their coding experience, primarily by providing quick answers, code templates, and ideas. However, the actual effectiveness in terms of problem-solving efficiency depends on the scenario: LLMs excel at some tasks (e.g. writing boilerplate or illustrating API usage) but might not reliably improve outcomes for complex problem-solving without user oversight. Users appreciate the convenience and support, but they also recognize that these models are not infallible. This balance of high perceived value with the need for caution is a recurring theme in the literature.

\subsection{LLMs in Software Development}

In the context of software development, studies have highlighted both the benefits and risks associated with LLM-assisted programming. GitHub Copilot, for instance, has been shown to improve developer efficiency but also introduce security vulnerabilities when users fail to critically evaluate generated code \cite{Ziegler2023, Sandoval2023}. Vaithilingam et al. \cite{Vaithilingam2022} explored user expectations versus actual experiences with LLMs, reporting discrepancies in trust and reliability, especially when handling complex logic or multi-step programming tasks. Nam et al. \cite{Nam2024} examined how developers use LLMs for code understanding, highlighting their role in assisting with unfamiliar programming constructs.

From an educational perspective, LLMs have been integrated into coding courses, with mixed results regarding their pedagogical impact. Haindl and Weinberger \cite{Haindl2024} reported that students appreciated the convenience of ChatGPT for explaining programming concepts but expressed concerns about its tendency to produce plausible yet incorrect explanations. Similarly, Huang et al. \cite{Huang2024} found that while LLMs helped students complete assignments more efficiently, they sometimes hindered deeper learning when used as a shortcut rather than a complement to traditional learning materials. Kazemitabaar et al. \cite{Kazemitabaar2023b} observed that novice programmers often relied on LLMs for completing introductory coding tasks but faced difficulties debugging incorrect outputs.

\subsection{Interaction Strategies and Prompt Engineering}

LLMs function as interactive problem-solving partners, adapting to user queries in real time. Ross et al. \cite{Ross2023} introduced the concept of the "programmer's assistant," emphasizing how conversational LLMs facilitate iterative debugging and code refinement through dialogue. Tang et al. \cite{Tang2024} analyzed how developers validate and repair LLM-generated code, demonstrating the importance of active engagement strategies such as rephrasing prompts and cross-referencing outputs with trusted sources. Barke et al. \cite{Barke2023} found that programmers iteratively refine their prompts to improve generated responses, often supplementing LLM output with alternative verification methods.

\subsection{Challenges and Limitations of LLMs in Coding}

While LLMs present a promising shift in programming assistance, their limitations necessitate further investigation. Issues such as hallucinations, lack of contextual memory, and ethical concerns regarding intellectual property continue to shape discussions on their adoption in professional and educational settings. Sandoval et al. \cite{Sandoval2023} warned of the security risks posed by LLM-generated code, as users may inadvertently introduce vulnerabilities without sufficient validation. 

The most prominent concern is that LLMs do not guarantee correctness. They can and do produce code with bugs, inefficiencies, or even logic errors. Multiple studies have documented cases where the AI’s output was subtly wrong or suboptimal, and an unalert user might not catch it. For example, a comparative study found that ChatGPT often gave “wrong or expired answers” for certain technical queries \cite{Liu2023ChatAssistComparison}. Copilot-generated code was observed to be less compact and sometimes lower in quality compared to human-written code \cite{Tang2024}. If users trust the AI too much, they might incorporate faulty code. The conversational format can be double-edged here: because LLMs can explain their reasoning in a confident, authoritative tone, users may be lulled into a false sense of security. Without careful testing or review, AI-suggested code can introduce bugs. Thus, the need for human verification remains a critical limitation – the AI is a helpful collaborator, but not a flawless one. To this end, Vaithilingam et al. \cite{Vaithilingam2022} also noted that user frustration arises when LLMs generate plausible but incorrect solutions, requiring additional debugging efforts. This study builds on prior research by providing empirical data on how programmers navigate these challenges, contributing to the broader literature on the use of LLM-driven conversational interfaces in programming.

\section{Method}
\subsection{Survey Design}

To understand how programmers interact with LLM-based coding assistants, we designed a survey consisting of both closed-ended and open-ended questions. The survey aimed to capture user experiences, interaction strategies, perceived benefits and limitations, and expectations for improvements in LLM-based coding support. The survey was structured into four main sections:

\subsubsection{LLM Usage and Effectiveness}
The first section focused on participants' frequency of LLM use and the perceived usefulness of these tools in coding tasks. Participants were asked:
\begin{itemize}
    \item How frequently they used LLMs for coding tasks in the past two weeks.
    \item Reasons for not using an LLM, if applicable.
    \item Alternative resources (e.g., documentation, lecture videos) they relied on for coding assistance.
    \item Which LLM(s) they used and the features they found most useful.
    \item Specific coding tasks or challenges (e.g., debugging, writing functions) where LLMs were applied and whether they were helpful.
\end{itemize}

\subsubsection{Interaction Strategies and Perceptions}
To gain insights into how users interact with LLMs, this section explored:
\begin{itemize}
    \item Strategies used when formulating prompts and refining responses.
    \item The most helpful aspects of LLMs in coding workflows.
    \item The biggest strengths and limitations of LLMs, along with suggestions for improvement.
\end{itemize}

\subsubsection{Design Expectations for Custom LLM Coding Assistants}
To inform the development of a custom LLM-integrated coding assistant, we asked participants:
\begin{itemize}
    \item What features they would like to see in a custom LLM for coding.
    \item Their preferences regarding the appearance of the assistant (e.g., humanoid avatar, robot avatar, simple icon).
    \item Factors that would make the tool more effective, efficient, and user-friendly.
\end{itemize}

\subsubsection{Final Open-Ended Reflections}
The survey concluded with an open-ended section where participants could provide additional feedback, ensuring we captured insights beyond predefined categories.


\subsection{Participants}
The survey was conducted with computer science students enrolled in a game development course at a large R1 university in the US, focusing on programming game mechanics using the C\# programming language in the Unity 3D game engine. The sample included a total of 143 student developers (39 females, 104 males) at the undergraduate and graduate level. The average age of students was 21.1 years old.


\subsection{Data Analysis}
We analyzed the open-ended responses to survey questions following qualitative thematic analysis. Specifically, following constant comparative analysis procedures~\cite{creswell2016qualitative}, we analyzed open-ended responses in three stages: 

\textbf{Open Coding:} We first performed open coding to identify recurring concepts and emerging themes in participants’ responses regarding their experiences, challenges, and expectations with LLMs for coding.

\textbf{Axial Coding:} Next, we grouped related themes into broader categories, such as strategies for interacting with LLMs, perceived strengths and weaknesses, and desired design improvements.

\textbf{Selective Coding:} Finally, we refined these categories into higher-level themes, allowing us to synthesize the findings and derive key design implications for LLM-integrated coding assistants.

To ensure rigor and consistency, all authors independently coded a subset of responses and discussed discrepancies to reach consensus on thematic categories. This iterative process allowed us to capture both common patterns and outlier perspectives, leading to a comprehensive understanding of how programmers interact with LLM-based coding tools.

\section{Results}

\subsection{LLM Usage for Coding Tasks}
Results revealed varying levels of engagement with LLMs for coding tasks. A minority of participants (7\%, n = 10) reported using LLMs daily, while 31\% (n = 45) engaged with them several times a week. The most common usage pattern was occasional interaction, with 40\% (n = 58) indicating they used LLMs once or twice a week. Notably, 22\% of the respondents (n = 31) reported never using LLMs for coding tasks.

These findings suggest that while LLMs have been integrated into many users’ workflows, their adoption remains inconsistent. The relatively low percentage of daily users may indicate that LLMs are seen as supplementary tools rather than primary resources for coding tasks. The next sections will explore the factors influencing these usage patterns, alternative resources utilized, and user perceptions of LLM effectiveness.

\subsection{Reasons for Not Using LLMs in Coding Tasks}

Participants provided a variety of reasons for not using LLMs in their coding workflows. The responses revealed six main themes: (1) reliance on alternative resources, (2) preference for independent learning, (3) lack of need or relevance, (4) concerns about learning and skill development, (5) trust and reliability issues, and (6) ethical and philosophical concerns.

\subsubsection{Reliance on Alternative Resources}
Many participants reported that existing resources, such as official documentation, lecture videos, and online forums, provided sufficient guidance for their coding needs:

\textit{"Unity documentation and videos went in-depth enough, and they were easy to follow."}  

\textit{"I figured I could find the answers to my questions on Stack Overflow or lectures."}

This suggests that some users perceive LLMs as redundant when high-quality instructional materials are readily available.

\subsubsection{Preference for Independent Learning}
Several participants indicated that they prefer to solve problems on their own, believing that this leads to better retention and deeper understanding:

\textit{"I prefer to figure things out on my own. It allows me to remember solutions better and help others in the future, even though it takes me more time and effort in the moment."}

\textit{"I believe the best way to learn programming is to actually write the code yourself and work through the bugs."}

These responses suggest that some programmers view struggle and hands-on problem-solving as essential components of the learning process.

\subsubsection{Lack of Need or Relevance}
Some participants simply did not feel the need to use LLMs, either because their assignments were straightforward or because LLMs did not naturally fit into their workflow:

\textit{"The coding portions were simple enough that I did not require an LLM for the code to work."}

\textit{"I never felt any need to, it didn't cross my mind to use one for any of this work."}

This implies that LLM adoption is often task-dependent, with users opting for LLMs only when they encounter challenges that other resources cannot address efficiently.

\subsubsection{Concerns About Learning and Skill Development}
A recurring concern was that relying on LLMs might hinder the learning process by encouraging dependency rather than skill-building:

\textit{"Using LLMs for coding, if you don't already have a grasp on that subject, means you aren’t actually learning anything, which will then cause you to be reliant on it."}

\textit{"I feel like I don't really learn if I ask LLMs to look through my code for me since it is a skill to spot things that may be wrong and to recognize how code segments operate."}

These responses highlight skepticism about whether LLMs foster genuine learning or merely facilitate shortcuts.

\subsubsection{Trust and Reliability Issues}
Some participants expressed concerns about the accuracy and trustworthiness of LLM-generated responses, preferring to rely on more validated sources.

\textit{"I don't trust AI with any technical information."}

\textit{"I think ChatGPT is useful, but if I have a specific question, I would rather find the answer in documentation or from a forum of actual developers than rely on it since it can be incorrect."}

This skepticism aligns with broader discussions about LLM hallucinations and the need for fact-checking AI-generated content.

\subsubsection{Ethical and Philosophical Concerns}
A smaller subset of participants rejected LLMs due to ethical, environmental, or ideological reasons:

\textit{"I don't use AI. It's horrible for the environment, and I never retain any of the information, so I don’t think it’s a good learning tool for me."}

\textit{"Academic integrity – not using one lets you understand your code more."}

These responses suggest that beyond usability concerns, some users actively resist AI adoption due to broader ethical considerations.


\subsection{CUIs for LLMs}

Participants were asked which LLM interfaces they used for coding tasks. The responses indicate that \textbf{ChatGPT} was the most popular CUI among the participants, with 90\% of respondents (n = 102) reporting its use. This suggests a strong preference for ChatGPT as an AI-assisted coding tool, likely due to its accessibility, conversational interface, and perceived effectiveness.

\textbf{GitHub Copilot} was the second most popular choice, with 26\% of respondents (n = 29) reporting its use. As an AI-powered code completion tool integrated into development environments, Copilot’s adoption reflects its growing role in assisting developers with writing and debugging code.

Other LLMs were used less frequently. \textbf{Claude} was utilized by 12\% (n = 14), followed by \textbf{DeepSeek} (4\%, n = 5 respondents), \textbf{Perplexity} (4\%, n = 4 respondents), and \textbf{Gemini} (3\%, n = 3). \textbf{Cursor}, a lesser-known tool, was mentioned by only one participant.

These findings highlight that while ChatGPT remains dominant, some participants explore alternative models based on their coding needs and preferences. The relatively lower usage of LLMs beyond ChatGPT and Copilot suggests that either awareness or accessibility of these models may be limited, or that their functionality does not yet match the expectations set by the more widely adopted tools. Future research could examine the contexts in which users select different LLMs and how specific model features influence programming workflows.

\subsection{Purposes of LLM Usage}

Participants reported various purposes for using LLMs in coding tasks over the past month. The most common purpose was \textbf{debugging}, with 73\% (82) of respondents using LLMs to identify and resolve errors in their code. \textbf{Syntax explanations} were also frequently sought, with 54\% (61) of participants relying on LLMs to clarify programming concepts and language constructs. Additionally, 47\% (53) of respondents used LLMs for \textbf{problem-solving strategies}, indicating that users leverage these models not only for direct assistance but also for conceptual guidance in tackling coding challenges. \textbf{Code generation} was a less frequent use case, with 22\% (25) of participants employing LLMs to generate code snippets, suggesting that users primarily see LLMs as tools for refining and understanding code rather than writing it from scratch. Only 2\% (2) of respondents reported using LLMs for other unspecified purposes. These findings indicate that while LLMs play a crucial role in debugging and learning, their use for direct code generation remains secondary for many users.

\subsection{Most Helpful Aspects of LLMs}

Participants identified several aspects of LLMs that were most helpful in their coding tasks. The most frequently cited benefit was \textbf{getting immediate feedback or assistance}, reported by 66\% (75) of respondents. This suggests that real-time interaction and responsiveness are key advantages of LLMs in programming workflows. Similarly, \textbf{clarity of explanations} was valued by 65\% (73) of participants, indicating that users rely on LLMs not just for providing solutions but also for improving their understanding of programming concepts. Additionally, 46\% (52) of respondents found \textbf{providing examples or templates} to be particularly useful, highlighting the role of LLMs in offering concrete reference points for problem-solving. Lastly, \textbf{speed of generating solutions} was appreciated by 24\% (27) of participants, suggesting that while rapid response is beneficial, users prioritize the accuracy, clarity, and interactive nature of LLM assistance over pure speed. These findings reinforce the idea that LLMs serve not just as code generators but as interactive learning aids that enhance comprehension and problem-solving efficiency.

\subsection{Perceived Usefulness of LLMs for Coding}

Participants reported varying levels of perceived usefulness regarding LLMs in assisting with their coding tasks. The majority of respondents found LLMs to be at least moderately useful, with 38\% (43) rating them as \textbf{moderately useful}, 25\% (28) as \textbf{very useful}, and 11\% (12) as \textbf{extremely useful}. This indicates that 74\% of participants perceived LLMs as significantly beneficial in their coding workflows. 

However, a subset of users expressed lower satisfaction with LLM assistance. \textbf{Slightly useful} was selected by 24\% (27) of respondents, suggesting that while LLMs provided some help, they may not have consistently met expectations or provided fully reliable solutions. A small fraction, 2\% (2) of respondents, found LLMs \textbf{not at all useful}, indicating that for some users, these tools did not offer meaningful support or were perceived as ineffective.

These results suggest that while LLMs are generally seen as valuable tools for coding, their perceived effectiveness varies based on user needs, expectations, and potentially the complexity of the coding tasks. Future research could explore the specific factors that influence LLM effectiveness, such as user expertise, task difficulty, or interaction strategies.

\subsection{Specific Use Cases of LLMs for Coding Tasks}

Participants were asked about the specific tasks or challenges for which they used LLMs and reported a range of tasks, yielding six primary themes: debugging, function explanations, best practices, software development-specific inquiries, code generation, and performance optimization/concept understanding.

\subsubsection{Debugging Assistance}
The most frequently reported use case for LLMs was debugging code and identifying errors. Participants reported using LLMs to troubleshoot issues ranging from syntax mistakes to logical errors in their code. LLMs were deemed particularly useful for pinpointing issues in specific functions, understanding error messages, and proposing fixes, as illustrated by the following quote from one of the respondents:

"\textit{Mostly debugging errors—especially when I overlook simple logic. Sometimes, I need help with syntax, and it helps me fill in the blanks.}"

However, some users noted that while LLMs provided quick solutions, they were sometimes generic or suboptimal, requiring further verification and refinement:

"\textit{I solved a small error in my logic through LLM. But when I described my results to LLM, it often couldn't combine my code well to find the cause of the error.}"

\subsubsection{Function Explanations and API Understanding}
Another prominent theme was using LLMs to understand functions, APIs, and programming concepts. Rather than consulting lengthy documentation, participants preferred LLMs for concise, interactive explanations of programming functions and language constructs:

"\textit{I primarily use LLMs to quickly look up definitions for various functions and to compare their use cases.}"

"\textit{I asked ChatGPT for detailed explanations about functions that I'm not sure about, specifically RotateAround and LookAt, and I also asked for a detailed explanation about Time.deltaTime.}"

Many found LLMs effective in breaking down complex topics into simpler explanations, but some noted that LLMs occasionally provided misleading or incorrect information, necessitating further validation:

"\textit{The provided code occasionally doesn't fully meet my expectations, especially as my requirements become more complex. In such cases, I view it more as a reference rather than a complete solution.}"

\subsubsection{Best Practices and Alternative Approaches}
Participants also turned to LLMs for best coding practices, optimization strategies, and alternative approaches to solving problems. They sought guidance on refactoring code for readability, improving performance, and structuring scripts effectively.

"\textit{To check the syntax and, upon completing the task, I was curious if there were any better or alternative ways to do it.}"

"\textit{It did help me in my script as I asked if there was another way it would approach the problem, and it reminded me to follow best practices and write helper functions.}"

While LLMs often provided useful recommendations, some respondents found that they needed to critically evaluate the suggestions, as the models sometimes overlooked context-specific nuances:

"\textit{I used an LLM for debugging, but often, I had to guide it by pointing out the inefficiencies in its responses before getting a satisfactory solution.}"

\subsubsection{Software Development-Specific Inquiries}
Beyond debugging and explanations, participants frequently relied on LLMs for software development-specific tasks, such as understanding testing strategies, version control, and input handling.

"\textit{I used an LLM to create tests for my function to make sure that my function is handling edge cases properly.}"

"\textit{Finding useful methods without needing to look through entire documentation.}"

"\textit{It helped me research how to make the application controller-friendly and map user input effectively.}"

This suggests that LLMs are used as on-demand programming assistants, streamlining workflows by quickly retrieving relevant information.

\subsubsection{Code Generation}
Some participants reported using LLMs to generate code snippets, reducing the need for manual implementation of repetitive patterns. LLMs were found helpful for writing function templates, generating boilerplate code, and suggesting implementations for known algorithms.

"\textit{Generating codes for a solved problem (e.g., A pathfinding, a basic movement controller).}"

"\textit{I used LLM to help me write comments for my codes and find the RGB for 'orange' color.}"

However, while LLMs were useful for quickly drafting code, some respondents found that the generated snippets often required heavy modification:

"\textit{I used an LLM to generate boilerplate code, but I had to modify most of it because it didn’t align with my project’s structure.}"

"\textit{The generated code worked but wasn't efficient, so I had to optimize and refactor a significant portion before it was usable.}"

\subsubsection{Performance Optimization and Concept Understanding}
A smaller but notable group of participants (approximately 14.7\% ) reported using LLMs to optimize performance, improve efficiency, and deepen their conceptual understanding.

"\textit{I used it for help understanding how to optimize my logic and organize code better.}"

"\textit{Asking syntax questions such as why there is an 'f' after setting values for variables, and it explained that it was to specify float values.}"

"\textit{I use LLMs to help me learn new languages, which makes the transition easier as I am unfamiliar with the syntax.}"

This suggests that LLMs serve not just as debugging assistants or code generators but also as learning tools for refining programming skills.

\subsection{Prompt Engineering Strategies for Interacting with LLMs}

Participants reported employing various strategies when interacting with LLMs to improve the quality of assistance they received. The analysis revealed six primary themes: 
(1) providing clear and specific prompts, (2) breaking tasks into smaller steps, (3) iterative refinement of prompts, (4) asking for explanations and justifications, (5) comparing multiple outputs, and (6) using LLMs as a learning tool.

\subsubsection{Providing Clear and Specific Prompts}
Many participants emphasized the importance of writing precise and well-defined prompts to obtain relevant and accurate responses from LLMs. Participants found that when they were explicit about their requirements, the model provided more useful answers. 

\textit{"Being specific with what was going wrong and giving relevant details helped the LLM provide more accurate responses."}

\textit{"I ask for specific deliverables from the LLM to ensure it gives me the exact format or information I need."}

This suggests that specificity in prompts enhances the effectiveness of LLMs by reducing ambiguity.

\subsubsection{Breaking Tasks into Smaller Steps}
A common strategy was breaking complex queries into smaller, manageable components before engaging with an LLM. Participants found that decomposing problems resulted in clearer and more actionable responses:

\textit{"I tried to describe my context and the problem in smaller steps, and this made the responses easier to apply to my code."}

\textit{"When I asked one big question, I often got an unclear response. Splitting my question into smaller ones made it easier to get helpful answers."}

This approach aligns with best practices in problem-solving, where structured inquiries lead to more effective troubleshooting and implementation.

\subsubsection{Iterative Refinement of Prompts}
Several participants noted that they adjusted or rephrased their prompts based on initial responses. This iterative interaction helped in refining the accuracy and relevance of the LLM's output.

\textit{"Sometimes the first answer isn’t quite right, so I rephrase my question and tweak it until I get something that makes sense."}

\textit{"I would take the initial response and adjust my wording to get a better or more targeted answer."}

This suggests that users engage in a feedback loop with LLMs, gradually improving the quality of assistance through prompt optimization.

\subsubsection{Asking for Explanations and Justifications}
Many users employed LLMs not just for direct answers but also for explanations and reasoning behind certain programming concepts or solutions. This helped them verify correctness and understand underlying principles.

\textit{"I make sure to ask ‘why’ when I get a response, so I can understand how it works instead of just copying the answer."}

\textit{"Rather than just getting a function, I ask the LLM to explain why this approach works and what alternative methods exist."}

This highlights the role of LLMs as educational tools that facilitate deeper conceptual understanding.

\subsubsection{Comparing Multiple Outputs}
Some participants deliberately requested different approaches to the same problem and then compared the outputs to assess which solution was most appropriate.

\textit{"I would ask the LLM for multiple ways to solve a problem and compare them to see which one makes the most sense."}

\textit{"I like to generate two or three different options and choose the best one based on my project’s needs."}

These quotes illustrate that by leveraging LLMs for multiple perspectives, developers could evaluate the trade-offs between different solutions.

\subsubsection{Using LLMs as a Learning Tool}
A subset of participants engaged with LLMs as an instructional resource rather than just a problem-solving assistant. They used the tool to reinforce learning, understand concepts better, and bridge knowledge gaps.

\textit{"I use LLMs to help me learn new programming concepts instead of just looking up solutions."}

\textit{"Sometimes I use it to clarify things I didn’t fully understand from the documentation or lectures."}

This indicates that some users treat LLMs as interactive tutors, aiding in self-directed learning.

\subsection{Strengths and Limitations of LLMs for Coding Tasks}

Participants identified both advantages and challenges associated with using LLMs for coding tasks. The analysis revealed three primary themes: (1) strengths, highlighting the benefits of LLMs in software development, (2) limitations, addressing key concerns and shortcomings, and (3) suggested improvements to enhance their usability.

\subsubsection{Strengths of LLMs for Coding Tasks}
Participants overwhelmingly cited the \textbf{efficiency} of LLMs as their greatest strength. Many users noted that LLMs significantly reduced the time spent searching for solutions and debugging errors:

\textit{"Strength is how much time it saves, and how it allows me to work faster without needing to search through multiple documentation pages."}

\textit{"The biggest strength is the speed and accessibility—rather than looking things up in different places, I can get quick answers in one interaction."}

\textbf{Accessibility and ease of use} were also emphasized, with participants appreciating how LLMs functioned as an on-demand assistant, reducing the need for extensive research:

\textit{"I think the biggest strength is the efficiency and ease of getting an immediate response."}

\textit{"Biggest strength is the quick explanation of things that would take much longer to find elsewhere."}

Additionally, some participants highlighted the role of LLMs in \textbf{learning and conceptual understanding}, using them as teaching tools.

\textit{"I think the strengths are that it's helpful in explaining concepts efficiently and clearly, which helps in learning new things."}

These insights suggest that LLMs serve as valuable time-saving and instructional resources for programmers.

\subsubsection{Limitations of LLMs for Coding Tasks}
Despite their strengths, LLMs exhibit several key limitations. The most frequently cited issue was \textbf{incorrect or misleading responses}. Participants expressed concerns about the reliability of the information provided, particularly when dealing with complex or nuanced coding problems:

\textit{"LLMs sometimes provide incorrect solutions, and if I don’t double-check, it can lead to more debugging later."}

\textit{"The biggest limitation is that it can hallucinate answers, making things up that sound plausible but are completely wrong."}

Another commonly reported issue was the \textbf{lack of context awareness}. Participants noted that LLMs struggle to fully understand the broader structure of their projects, leading to incomplete or generic recommendations.

\textit{"LLMs don't have full knowledge of my codebase, so sometimes the suggestions don’t fit the existing structure of my project."}

\textbf{Lack of reasoning and adaptability} was also mentioned, with some users stating that LLMs struggle with complex problem-solving beyond basic syntax and logic errors:

\textit{"It can give you an answer, but it doesn’t always think critically about the problem—it just outputs something that looks right."}

These findings highlight the importance of validating LLM-generated responses before implementation.

\subsubsection{Suggested Improvements}
Participants provided several suggestions on how LLMs could be improved to better meet their needs. The most frequent recommendation was \textbf{enhancing contextual awareness}, allowing LLMs to retain information across interactions:

\textit{"If LLMs could remember previous responses and provide continuity in conversations, that would be a game-changer."}

Users also suggested \textbf{better handling of uncertainty}, where LLMs should acknowledge their limitations instead of confidently providing incorrect answers:

\textit{"LLMs should indicate when they are unsure instead of making things up—it would save a lot of time."}

Additionally, some participants proposed improvements in \textbf{code customization and adaptability}, such as allowing LLMs to better align responses with specific coding styles and project structures.

\textit{"It would be great if I could set some parameters so the LLM would generate code in my preferred format or style."}

\subsection{Expectations from Custom LLMs for Coding}
Participants were also asked about their expectations from a custom LLM-integrated coding assistants and provided several feature suggestions. The analysis revealed six primary themes: (1) improved prompting and guidance, (2) context awareness and memory, (3) more precise and actionable responses, (4) multimodal capabilities, (5) code customization and adaptability, and (6) debugging and error explanations.

\subsubsection{Improved Prompting and Guidance}
Several participants expressed interest in having better guidance on how to interact with the LLM effectively. They suggested features that would assist users in crafting clearer prompts and improving query formulation:

\textit{"Step-by-step guidance about how to navigate [the] API and how to structure questions to get the best response."}

\textit{"How to edit the prompt to make the LLM answer more relevant to my specific issue would be very useful."}

This suggests a need for built-in mechanisms to help users refine their interactions with the assistant.

\subsubsection{Context Awareness and Memory}
A frequent request was for the LLM to retain context across interactions, allowing it to remember previous discussions and project-specific details:

\textit{"It can understand the hierarchy of a Unity project and remember previous responses to build upon them."}

\textit{"More ways to input context so I don’t have to repeat information in every query."}

This highlights the importance of persistent memory and improved contextual understanding.

\subsubsection{More Precise and Actionable Responses}
Participants wanted responses that were more direct, actionable, and aligned with their specific coding issues.

\textit{"I would like to be able to get exact hints as opposed to general explanations that may or may not be useful."}

\textit{"More relevant responses—sometimes the LLM suggests things that don’t apply to my case at all."}

These findings indicate that improving response precision would enhance the usefulness of the LLM.

\subsubsection{Multimodal Capabilities (Images, Diagrams, Voice)}
Some users suggested incorporating visual aids, such as images and diagrams, to complement text-based responses:

\textit{"I'd like to see some images instead of just text explanations—maybe flowcharts or UML diagrams for complex logic."}

\textit{"Response in image and text combination for better understanding."}

This suggests that adding multimodal capabilities could improve comprehension and engagement.

\subsubsection{Code Customization and Adaptability}
A subset of users requested features that allow the LLM to align its responses with specific coding styles, preferences, and project needs:

\textit{"I would love a way to set preferences so that generated code follows my style, including naming conventions and formatting."}

\textit{"Let me define coding patterns or structure so that responses are more consistent with my existing codebase."}

This highlights the need for customizable outputs tailored to individual developer preferences.

\subsubsection{Debugging and Error Explanations}
Many participants wanted better debugging assistance, including more insightful explanations of errors and step-by-step fixes:

\textit{"A feature that explains errors in detail instead of just suggesting generic fixes would be very useful."}

\textit{"It would be great if the LLM could analyze my error messages and tell me not just how to fix them but why they occurred."}

These responses emphasize the importance of enhancing LLMs' debugging capabilities.

\subsubsection{Clear Instructions and Guidance}
Many participants emphasized the importance of structured guidance on how to use the tool effectively. They suggested improvements in user onboarding, documentation, and built-in prompts.

\textit{"Clear instructions and organized requirements on how to phrase queries would make this tool much easier to use."}

\textit{"There should be a variety of options for how it presents information, so new users can navigate it more effectively."}

Providing structured guidance can help users formulate better queries, improving the overall experience.

\subsubsection{Interactive and Real-Time Feedback}
A subset of participants requested features that enable real-time guidance and interactive debugging support.

\textit{"Step-by-step debugging walkthroughs with live feedback would make this tool incredibly useful."}

\textit{"Integration with IDEs for real-time assistance and corrections while coding."}

Interactive support mechanisms could enhance usability by providing immediate and iterative guidance.

\subsubsection{Appearance Preferences}

Participants expressed varying preferences regarding the visual representation of the custom LLM-integrated coding assistant. The majority favored a \textbf{simple icon or minimalist UI}, emphasizing that a non-intrusive design would enhance usability. 

\textit{"I like the simple icon."}  
\textit{"A clean interface without unnecessary visuals is ideal."}  

Some preferred a \textbf{robotic or AI-themed avatar}, aligning with the perception of an AI assistant.

\textit{"A robot avatar would fit the AI theme well."}

A smaller group suggested a \textbf{humanoid avatar} for a more engaging experience, while others favored \textbf{voice interaction} for added accessibility.

\textit{"A friendly humanoid avatar might make interactions feel more natural."}  
\textit{"Having voice functionality would make it feel more interactive."}  

Additionally, several users requested \textbf{customization options} to allow them to select between different visual styles:

\textit{"Let users choose from different appearances—simple, robot, or human-like."}

These insights suggest that a customizable UI with a default minimalistic design could best accommodate diverse user preferences.

\section{Discussion}


This study investigated how student developers interact with LLM-based coding assistants, their perceived strengths and limitations, and their expectations for future improvements. Our findings highlight key insights into how CUIs for LLM-based coding can be designed to better support coding workflows.

One of the most notable insights from our study is that programmers primarily use LLMs for debugging, syntax explanations, and problem-solving rather than direct code generation. This contrasts with prior work that has emphasized LLMs as tools for code automation~\cite{Liu2023ChatAssistComparison}. Instead, our findings suggest that conversational AI serves as an interactive partner, helping users understand their code, troubleshoot issues, and refine their programming strategies. Additionally, we found that users do not passively accept LLM outputs but instead engage in an iterative refinement process, adjusting their prompts, comparing multiple solutions, and seeking explanations to validate responses. This aligns with prior research on prompt engineering strategies ~\cite{Ross2023}, but we extend this understanding by showing how these strategies shape the overall effectiveness of LLM-assisted programming.

Despite these benefits, users frequently expressed concerns regarding LLM response accuracy and reliability. Many participants reported that while LLMs provided useful starting points, the generated responses were sometimes incorrect or misleading, requiring additional validation. This aligns with studies highlighting the risks of AI hallucinations in programming contexts~\cite{Sandoval2023}. Additionally, a major frustration reported by users was the lack of context retention, requiring them to repeatedly provide the same background information in multi-turn interactions. This limitation has been widely noted in prior work on conversational AI ~\cite{Vaithilingam2022}, and our findings reinforce the need for context-aware LLMs that can maintain relevant session history.

Beyond these challenges, our study highlights an increasing demand for customization and integration within coding environments. Participants expressed a strong preference for being able to adjust the verbosity of responses, define their coding style preferences, and seamlessly integrate LLM assistance within IDEs and debugging tools. Additionally, several users suggested expanding the modality of interactions by incorporating visual aids such as syntax trees, UML diagrams, and interactive debugging visualizations to complement textual responses. These findings suggest that the future of CUIs for LLM-based coding assistance lies in adaptive, multimodal, and context-aware systems.

\subsection{Barriers to LLM Adoption for Coding}

While LLM-based coding assistants offer notable benefits, their adoption is not universal. Our findings reveal a range of barriers, both practical and conceptual, that influence whether programmers choose to engage with LLMs in their workflows. These barriers align with existing research on tool adoption, AI trust, and self-directed learning in programming. In this section, we discuss these findings in relation to prior work, highlighting the complex interplay between perceived usefulness, learning strategies, and ethical considerations.

Many participants cited existing resources such as documentation, lecture videos, and forums (e.g., Stack Overflow) as sufficient for their needs, suggesting that LLMs are often viewed as redundant. Prior research indicates that developers heavily rely on structured, authoritative sources when seeking help with coding tasks~\cite{treude2019documentation, pimentel2019stackoverflow}. Unlike LLMs, which generate answers dynamically, these resources provide stable, curated, and community-validated information, which may explain why some users perceive them as more reliable.

Moreover, documentation and human-curated resources foster a deeper engagement with the underlying concepts rather than presenting immediate solutions. This preference aligns with literature on learning scaffolding, which suggests that structured instructional materials support long-term retention better than AI-generated answers~\cite{koedinger2013learning}. As such, while LLMs may offer efficiency gains, they may not always align with users' preferred learning strategies, particularly among those who value deep engagement over immediate answers.

A recurring concern among participants was that using LLMs might hinder the learning process, making users reliant on AI-generated solutions rather than engaging in problem-solving. This aligns with the concept of cognitive offloading, where reliance on external tools reduces the need for internal processing~\cite{norman1993cognitive}. While cognitive offloading can enhance efficiency in professional settings, research suggests that it may be detrimental in educational contexts where skill development and conceptual understanding are critical~\cite{williams2022ai}.

Similar concerns have been raised in research on AI-assisted learning, where studies suggest that excessive reliance on AI-generated content can lead to surface-level understanding rather than conceptual mastery~\cite{kim2022learning}. While some work has suggested that AI-driven tools can enhance learning when used interactively~\cite{holstein2019designing}, our findings indicate that programmers perceive LLMs as more of a shortcut than an interactive learning tool. This distinction is important for designing AI-assisted learning experiences that balance efficiency with meaningful engagement.

In addition, trust remains a significant barrier to LLM adoption. Some participants expressed skepticism about the accuracy of AI-generated responses, preferring to rely on documented sources and community-vetted information. These concerns are consistent with prior research on LLM hallucinations, where AI-generated code can be syntactically correct but semantically flawed~\cite{bender2021dangers, Vaithilingam2022}. Developers working on critical software tasks, particularly those involving debugging or optimization, often cross-check AI-generated outputs due to these concerns~\cite{ziegler2023measuring}.

The issue of AI trust is further complicated by the opaque nature of LLM reasoning. Unlike human experts or documentation that cite sources and provide structured explanations, LLMs generate responses probabilistically, making it difficult to verify their reliability~\cite{Sandoval2023}. Prior research suggests that AI-assisted coding tools should incorporate confidence indicators, citations, and explanations of reasoning to enhance trust and usability~\cite{wang2022explainable}. Our findings reinforce this need, as participants repeatedly expressed a preference for sources they could verify over AI-generated responses that lacked transparency.

Another reason for avoiding LLMs was that participants did not perceive them as necessary. When coding tasks were straightforward or well-documented, LLMs did not provide enough additional value to justify their use. This aligns with broader research in HCI and software engineering showing that tool adoption is often task-dependent~\cite{grudin1994adoption}. Developers are more likely to adopt AI-driven tools when faced with complex, unfamiliar, or ambiguous coding challenges~\cite{fischer2001users}, rather than routine programming tasks where conventional resources suffice.

Moreover, studies on developer workflow integration highlight that new tools must provide seamless value addition to be widely adopted~\cite{ko2007exploratory}. If a tool does not meaningfully reduce cognitive effort or improve efficiency beyond existing workflows, users are unlikely to adopt it. Our findings suggest that the current generation of LLMs has yet to demonstrate sufficient advantage in well-structured coding tasks, leading many programmers to forgo their use entirely.

A subset of participants rejected LLMs for ethical and philosophical reasons, citing concerns about academic integrity, environmental impact, and AI’s role in creative fields. Ethical concerns about AI-generated content, including plagiarism and authorship attribution, have been widely discussed in educational technology and HCI research~\cite{bailey2022ethics}. Similarly, the environmental cost of training and running LLMs has raised concerns about the sustainability of large-scale AI models~\cite{strubell2019energy}.

Beyond environmental and academic integrity issues, some participants expressed a fundamental dislike of AI in such creative tasks as programming. Prior research suggests that AI skepticism often stems from concerns about automation displacing human expertise, particularly in domains that rely on creativity and problem-solving~\cite{dreyfus2007human}. Our findings suggest that this skepticism extends to programming, where some users prefer to develop solutions independently rather than rely on AI-generated outputs.

\subsubsection{Implications for LLM Design and Adoption}

Understanding these barriers to LLM adoption provides critical insights for improving AI-assisted coding tools. For LLMs to gain broader acceptance, they must:

\begin{itemize}
    \item \textbf{Complement, rather than replace, existing resources:} LLMs should be integrated alongside documentation and tutorials, helping users cross-reference information rather than providing standalone answers.
    \item \textbf{Support active learning strategies:} AI coding assistants should guide users through problem-solving rather than simply providing direct solutions, aligning with constructivist learning principles~\cite{jonassen1999constructivist}.
    \item \textbf{Increase transparency and reliability:} Implementing confidence indicators, multiple solution pathways, and citations to external sources could improve trust.
    \item \textbf{Allow customization for different learning styles:} Users should be able to adjust response verbosity, explanation depth, and reasoning transparency, catering to both novice and expert programmers.
    \item \textbf{Improve task relevance through adaptive assistance:} LLMs should better detect when users need assistance and proactively offer contextually relevant interventions rather than requiring users to initiate all interactions.
\end{itemize}

By addressing these concerns, LLM-based coding assistants can evolve into tools that not only provide immediate solutions but also support long-term skill development and trustworthy collaboration.

\subsection{Design Guidelines for Custom LLM Coding Assistants}

Based on our findings, we propose a set of design guidelines to enhance the usability, effectiveness, and trustworthiness of LLM-based coding assistants. These guidelines address key user concerns, including context awareness, interaction strategies, response reliability, multimodal support, and integration within software development environments.

\subsubsection{Context Awareness and Persistent Memory}

A major frustration among users was the need to restate context repeatedly when interacting with LLMs. Users expressed a strong preference for assistants that can retain session memory, allowing for a more natural and efficient conversation flow.

\textbf{Guidelines:}
\begin{itemize}
    \item Implement session-based memory to maintain relevant information throughout a coding session.
    \item Enable the assistant to track recently used variables, functions, and files to provide more contextually appropriate suggestions.
    \item Allow users to manually pin key details (e.g., project goals, function names) that persist across multiple queries.
\end{itemize}

\subsubsection{Transparency and Confidence Calibration}

Users reported concerns about LLMs providing misleadingly confident responses, even when incorrect. To improve trust, LLMs should communicate uncertainty more effectively and provide verification mechanisms.

\textbf{Guidelines:}
\begin{itemize}
    \item Display confidence scores for responses, especially when multiple solutions exist.
    \item Provide explanatory reasoning for suggested solutions, referencing documentation when possible.
    \item Allow users to toggle a "verification mode" where the assistant flags uncertain responses and suggests cross-checking with official sources.
\end{itemize}

\subsubsection{Adaptive Guidance for Prompt Refinement}

Users frequently engaged in iterative query refinement to improve the quality of LLM responses. An intelligent coding assistant should proactively assist users in structuring more effective prompts.

\textbf{Guidelines:}
\begin{itemize}
    \item Provide real-time suggestions for refining queries, such as breaking down complex requests into smaller parts.
    \item Offer an auto-complete system that suggests commonly used coding prompts based on previous interactions.
    \item Incorporate example-driven recommendations, where users can see sample prompts that have led to high-quality responses.
\end{itemize}

\subsubsection{Multimodal Support for Enhanced Understanding}

Text-based responses alone were sometimes insufficient for conveying complex programming concepts. Users expressed a desire for visual explanations, particularly for debugging and algorithmic reasoning.

\textbf{Guidelines:}
\begin{itemize}
    \item Integrate diagram generation (e.g., syntax trees, UML models, flowcharts) to illustrate coding concepts.
    \item Support annotated code explanations, where users can highlight parts of their code and receive targeted explanations.
    \item Allow voice-based interaction for hands-free querying and conversational debugging assistance.
\end{itemize}

\subsubsection{Seamless Integration with Development Environments}

Participants noted that switching between an LLM interface and their coding environment disrupted workflow. A well-designed assistant should be embedded within existing IDEs, version control systems, and debugging tools.

\textbf{Guidelines:}
\begin{itemize}
    \item Enable in-line assistance within IDEs, such as providing real-time suggestions while coding.
    \item Offer integration with Git workflows, allowing users to query commit history, detect regressions, and auto-generate commit messages.
    \item Support debugging assistance, where users can directly feed error logs into the assistant for targeted troubleshooting suggestions.
\end{itemize}

\subsubsection{Customizability and User Control}

Users have varying preferences regarding response verbosity, explanation depth, and coding style. Providing configurable settings allows users to tailor the assistant’s behavior to their needs.

\textbf{Guidelines:}
\begin{itemize}
    \item Allow users to set preferences for response length, detail level, and code formatting.
    \item Support different coding paradigms and frameworks, enabling users to specify preferred libraries or design patterns.
    \item Offer a learning mode where the assistant provides step-by-step guidance versus an expert mode for direct solutions.
\end{itemize}

\subsubsection{Validation and Error Handling Mechanisms}

Users highlighted the need for LLMs to assist not just in generating code but also in validating and refining it. To increase reliability, coding assistants should incorporate self-checking and debugging mechanisms.

\textbf{Guidelines:}
\begin{itemize}
    \item Enable automated code validation, where the assistant can check syntax and logic errors before suggesting a solution.
    \item Allow users to request alternative solutions, helping them explore multiple ways to solve a problem.
    \item Provide debugging walkthroughs, where the assistant explains potential reasons for errors step by step.
\end{itemize}

\subsection{Limitations and Future Work}

While this study provides valuable insights, several limitations must be acknowledged. First, our findings are based on self-reported survey data, which may introduce biases such as selective recall or overestimation of LLM effectiveness. Future studies could incorporate interaction log analysis to validate self-reported usage patterns. Second, our sample included computer science students taking a game programming course, but did not distinguish between novice and expert programmers in detail. A more granular analysis of experience levels could reveal nuanced differences in LLM interaction styles. 

Additionally, while our study focused on text-based LLMs, future research should explore the integration of multimodal AI assistants, including those supporting voice interactions and real-time debugging with visual explanations. Finally, as LLMs continue to evolve, longitudinal studies tracking user behavior over time will be crucial for understanding how programming workflows adapt to increasingly advanced AI capabilities.

Lastly, our findings suggest that while LLMs hold promise for assisting programmers, their adoption is influenced by learning preferences, trust, perceived task relevance, and ethical considerations. These barriers highlight the need for AI-driven tools that prioritize transparency, reliability, and pedagogical alignment. Future research should explore adaptive AI systems that balance efficiency with learning, as well as ethical frameworks for responsible AI integration in software development.

\section{Conclusion}

This study examined how programmers engage with LLM-based conversational assistants, providing insights into their perceived effectiveness, common usage strategies, and limitations. Our findings reveal that while LLMs are valued for debugging, syntax explanations, and problem-solving, concerns about contextual awareness and response accuracy remain. 

We proposed a set of design principles to inform the development of more effective custom LLM coding assistants, emphasizing context retention, transparency, multimodal support, and IDE integration. These insights contribute to ongoing discussions in human-computer interaction and conversational user interfaces by offering empirical evidence on the evolving role of AI-powered assistants in software development.

Future research should build on these findings by investigating longitudinal adoption trends, multimodal interaction strategies, and the role of AI in collaborative software engineering environments. By addressing these open questions, we can further refine LLM-based conversational interfaces to better support the needs of programmers and software developers.

\bibliographystyle{ACM-Reference-Format}
\bibliography{sample-base}

\end{document}